\documentclass[conference]{IEEEtran}
\IEEEoverridecommandlockouts
\usepackage{cite}
\usepackage{amsmath,amssymb,amsfonts}
\usepackage{graphicx}
\usepackage{textcomp}
\usepackage{xcolor}
\usepackage{balance}
\usepackage{cite}
\usepackage{hyperref}
\usepackage{amsmath,amssymb,amsfonts}
\usepackage{amsmath}
\usepackage{amsthm}
\DeclareMathOperator*{\argmax}{argmax}
\DeclareMathOperator*{\argmin}{argmin}
\usepackage{graphicx}
\usepackage{textcomp}
\usepackage{xcolor}
\usepackage{graphicx}
\usepackage{float}
\usepackage{subfigure}
\usepackage{amsmath}
\usepackage{amsfonts,amssymb}
\usepackage{mathrsfs}
\usepackage{mathtools}
\usepackage{bm}
\usepackage{multirow}
\usepackage{array}
\usepackage{amssymb}
\usepackage{amsmath}
\usepackage{cite}
\usepackage{url}
\usepackage{xcolor}
\usepackage{cite,graphicx,amsmath,amssymb}
\usepackage{subfigure}
\usepackage{fancyhdr}
\usepackage{mdwmath}
\usepackage{mdwtab}
\usepackage{caption}
\usepackage{amsthm}
\usepackage{setspace}
\usepackage{bm}
\usepackage{mathtools}
\usepackage{dsfont}
\usepackage{bbm}
\usepackage{algorithm}
\usepackage{algorithmic}

\newtheorem{theorem}{Theorem}

\newtheorem{lemma}{Lemma}

\newtheorem{corollary}{Corollary}

\makeatletter
\newcommand{\biggg}{\bBigg@{3}}
\newcommand{\Biggg}{\bBigg@{3.5}}
\makeatother
\def\BibTeX{{\rm B\kern-.05em{\sc i\kern-.025em b}\kern-.08em
    T\kern-.1667em\lower.7ex\hbox{E}\kern-.125emX}}
\begin{document}

\title{Joint Receive Antenna Selection and Beamforming in RIS-Aided MIMO Systems}

\author{\IEEEauthorblockN{Chongjun~Ouyang$^{\star}$, Ali~Bereyhi$^{\dag}$, Saba Asaad$^{\dag}$, Ralf~R.~M\"{u}ller$^{\dag}$, and Hongwen~Yang$^{\star}$}
$^{\star}$School of Information and Communication Engineering, Beijing University of Posts and Telecommunications\\
$^{\dag}$Institute for Digital Communications (IDC), Friedrich-Alexander-Universit\"{a}t Erlangen-N\"{u}rnberg\\
Email: $^{\star}$\{DragonAim, yanghong\}@bupt.edu.cn,~$^{\dag}$\{ali.bereyhi, saba.asaad, ralf.r.mueller\}@fau.de}

\maketitle

\begin{abstract}
This work studies a low-complexity design for reconfigurable intelligent surface (RIS)-aided multiuser multiple-input multiple-output systems. The base station (BS) applies receive antenna selection to connect a subset of its antennas to the available radio frequency chains. For this setting, the BS switching network, uplink precoders, and RIS phase-shifts are jointly designed, such that the uplink sum-rate is maximized. The principle design problem reduces to an NP-hard mixed-integer optimization. We hence invoke the weighted minimum mean squared error technique and the penalty dual decomposition method to develop a tractable iterative algorithm that approximates the optimal design effectively. Our numerical investigations verify the efficiency of the proposed algorithm and its superior performance as compared with the benchmark.
\end{abstract}


\section{Introduction}
Large multiple-input multiple-output (MIMO) technology, often referred to as massive MIMO, is known to boost the spectral efficiency of wireless channels. However, its fully-digital implementation with a dedicated radio frequency (RF) chain at each antenna suffers from expensive hardware costs and excessive energy consumption. Numerous potential approaches have been introduced to alleviate this issue over the past years; see \cite{Ouyang2020,Ali2018,Gershman2004} and the references therein. Among them is antenna selection that sets only a small subset of antennas active in each coherence time. This way it reduces the number of required RF chains, and hence the overall RF cost without significantly degrading the performance.

On a parallel track, recent advances in RF micro electromechanical systems have made programmable meta-surfaces a reality, among which reconfigurable intelligent surfaces (RISs) have received a growing attention \cite{Wu2020}. An RIS is a planar array comprising low-cost passive reconfigurable reflecting elements (REs). The main idea behind RIS is to tune the phase responses of REs, such that the propagation environment is smartly reconfigured. Deploying an RIS introduces additional degrees of freedom to the system and improves the wireless link performance, e.g., spectral efficiency \cite{Liu2021}.

Against this backdrop, it is natural to combine the two technologies of antenna selection and RIS to improve system performance while reducing the RF cost; see \cite{Wang2022,Abdullah2022,He2022,Xu2022,Rezaei2022} for some recent studies in this respect. The early attempt in \cite{Wang2022} studied the joint antenna selection and RIS phase-shifts design of a single-user multiple-input single-output (MISO) case. The result is extended to the single-user MIMO case in \cite{He2022}. Extensions to multiuser MISO, multiuser MIMO (MU-MIMO), and multi-cell settings are further discussed in \cite{Abdullah2022}, \cite{Xu2022}, and \cite{Rezaei2022}, respectively.
\subsection{Contributions}
Despite research progresses on this topic, the available literature is still restricted to some particular scenarios. These restrictions are mainly in two respects: firstly, these earlier studies only considered single-stream transmission and ignored the design of precoders at the user terminals (UTs); Secondly, most existing works, e.g., \cite{He2022,Xu2022,Rezaei2022}, assume that RE phase-shifts are continuously tuned. From the implementational viewpoint, this is an unrealistic assumption, as most proposed realizations of RISs apply quantized phase-shifts on the receive signal.

In this paper, we consider an RIS-aided MU-MIMO system, where each RE takes its phase-shift from a discrete set, and each UT sends multiple data streams to the base station (BS). We maximize the throughput of this system by proposing a joint design that optimizes the UT precoders, RIS phase-shifts, and the switching network at the BS. The basic design problem is an NP-hard problem dealing with integer programming. Starting from this problem, we propose an iterative algorithm through the following lines of contributions: 1) We propose a penalty dual decomposition (PDD)-based method to tackle the non-convex joint design problem via capitalizing on the weighted minimum mean squared error (WMMSE) technique. 2) To further alleviate the complexity, we next propose an alternative algorithm based on sequential optimization (SO). This algorithm is shown to reduce the complexity at the expense of performance losses. This result verifies the efficiency of our PDD-based algorithm and depicts the complexity-performance trade-off. We verify our low-complexity design through numerical experiments. Our numerical results demonstrate that the proposed algorithms significantly outperform the benchmark.
\subsection{Notation}
Throughout this paper, scalars, vectors, and matrices are denoted by non-bold, bold lower-case, and bold upper-case letters, respectively. For the matrix $\mathbf A$, $[\mathbf A]_{i,j}$, ${\mathbf{A}}^{\mathsf T}$, and ${\mathbf{A}}^{\mathsf H}$ denote the $(i,j)$th entry, transpose, and transpose conjugate of $\mathbf A$, respectively. For the square matrix $\mathbf B$, ${\mathbf B}^{\frac{1}{2}}$, ${\mathbf B}^{-1}$, ${\mathsf{tr}}(\mathbf B)$. and $\det(\mathbf B)$ denote the principal square root, inverse, trace, and determinant of $\mathbf B$, respectively. The notation $[\mathbf a]_{i}$ denotes the $i$th entry of vector $\mathbf a$, and $\mathsf{diag}\{\mathbf a\}$ returns a diagonal matrix whose diagonal elements are entries of $\mathbf a$. The identity matrix, zero matrix, and all-one vector are represented by $\mathbf I$, $\mathbf 0$, and $\mathbf 1$, respectively. The matrix inequalities ${\mathbf A}\succeq{\mathbf 0}$ and ${\mathbf A}\succ{\mathbf 0}$ imply that $\mathbf A$ is positive semi-definite and positive definiteness, respectively. The operation $\angle s$ extracts the phase of the complex value $s$. The set $\mathbbmss{C}$ stands for the complex plane and notation ${\mathbbmss{E}}[\cdot]$ represents mathematical expectation. The Hadamard product is shown by $\odot$, and $[K]$ represents the integer set $\{1, \ldots ,K\}$. Finally, ${\mathbf x}\sim{\mathcal{CN}}\left({\mathbf 0},{\mathbf X}\right)$ denotes a circularly symmetric complex Gaussian vector with mean zero and covariance matrix $\mathbf X$.


\begin{figure}[!t]
    \centering
    \includegraphics[width=0.25\textwidth]{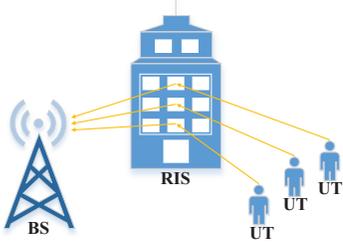}
    \caption{Illustration of a RIS-aided MU-MIMO system}
    \label{system_model}
\end{figure}

\section{System Model and Problem Formulation}
We consider uplink transmission in a RIS-aided multiuser MIMO setting in which $K$ multiple-antenna UTs send their encoded messages simultaneously to a BS with $N$ antennas. A schematic view of the setting is given in {\figurename} {\ref{system_model}}. The number of antennas at UT $k\in[K]$ is assumed to be $N_k$. The communication from all the UTs to the BS is aided by an RIS with a set of $M$ REs. It is assumed that the direct UT-to-BS links are blocked due to unfavorable propagation conditions.

UT $k\in[K]$ employs the linear precoder ${\mathbf P}_k\in{\mathbbmss{C}}^{N_k\times L_k}$ to construct its transmit signal ${\mathbf x}_k\in{\mathbbmss{C}}^{N_k}$ from its vector of symbols ${\mathbf s}_k\in{\mathbbmss C}^{L_k}$ that is considered to have zero mean and identity covariance matrix. We consider a standard multiple access channel (MAC) in which UTs send independent data streams, i.e., ${\mathbbmss E}\left[{\mathbf s}_k{\mathbf s}_{k'}^{\mathsf{H}}\right]={\mathbf 0}$ for $k\neq k'$. In this setting, the received signal at the BS is given by
{\setlength\abovedisplayskip{2pt}
\setlength\belowdisplayskip{2pt}
\begin{align}\label{System_Model_Received}
{\mathbf{y}}=\mathbf{G}{\bm\Phi}\sum\nolimits_{k=1}^{K}\mathbf{H}_k{\mathbf x}_k+\mathbf{n}.
\end{align}
}The terms appearing in \eqref{System_Model_Received} are defined as follows:
\begin{itemize}
  \item $\mathbf{G}\in{\mathbbmss{C}}^{N\times M}$ is the RIS-to-BS channel matrix.
  \item $\mathbf{H}_k\in{\mathbbmss{C}}^{M\times N_k}$ is the matrix of channel coefficients from UT $k$ to the RIS.
  \item $\mathbf{n}$ denotes additive white Gaussian noise (AWGN) with mean zero and variance $\sigma^2$, i.e., $\mathbf{n}\sim{\mathcal{CN}}\left(\mathbf{0},\sigma^2\mathbf{I}\right)$.
  \item ${\bm\Phi}\in{\mathbbmss{C}}^{M\times M}$ is the diagonal matrix of RIS phase-shifts, i.e., ${\bm\Phi}={\mathsf{diag}}\left\{{\bm\phi}\right\}$ for ${\bm\phi}=[\phi_{1},\ldots,\phi_{M}]^{\mathsf{T}}$, where $\phi_{m}={\rm{e}}^{{\rm{j}}\theta_{m}}$ for $m\in[M]$ with $\mathrm j$ denoting the imaginary unit and $\theta_{m}$ being the phase-shift applied by the $m$th RE of the RIS on its received signal.
\end{itemize}
We assume that REs apply quantized phase-shifts with $Q$ bits of quantization. This means that $\theta_m$ is adjusted to $2^Q$ discrete values. The discrete levels are specified by uniform quantization of the phase interval $[0, 2\pi)$, i.e.,
{\setlength\abovedisplayskip{2pt}
\setlength\belowdisplayskip{2pt}
\begin{align}
\phi_m\in{\mathcal{F}}_Q=\left\{{\rm e}^{{\rm j}2\pi\frac{t-1}{2^{Q}}}:t\in\left[2^Q\right]\right\}.
\end{align}
}The system operates in the time division duplexing (TDD) mode. The channel state information (CSI) is hence estimated in the uplink training phase via pilot sequences. We assume that the pilots are mutually orthogonal and that the estimation error is negligible. The BS thus learns perfectly the CSI. Details on channel estimation for RIS-aided MIMO systems are found in \cite{Swindlehurst2022} and the references therein.
\subsection{Receive Antenna Selection}
The BS has $T<N$ RF chains and hence uses a switching network to select a subset of receive
antennas. This switching network connects the selected antennas to the available $T$ RF chains at the BS. As a result, the received signal at the RF front-end of the BS is given by
{\setlength\abovedisplayskip{2pt}
\setlength\belowdisplayskip{2pt}
\begin{align}
{\mathbf r}={\mathbf S}{\mathbf y}={\mathbf S}\mathbf{G}{\bm\Phi}\sum\nolimits_{k=1}^{K}\mathbf{H}_k{\mathbf x}_k+{\mathbf S}{\mathbf n},
\end{align}
}where ${\mathbf S}={\{0,1\}}^{T\times N}$ is the antenna selection matrix with
{\setlength\abovedisplayskip{2pt}
\setlength\belowdisplayskip{2pt}
\begin{align}
[{\mathbf S}]_{t,n}=\begin{cases}
1& {\text{antenna}}~n~{\text{is connected to}}~{\text{RF chain}}~t\\
0& {\text{otherwise}}
\end{cases}.\nonumber
\end{align}
}Since $\mathbf S$ is a fat permutation matrix, we have ${\mathbf{S}}{\mathbf{S}}^{{\mathsf{H}}}={\mathbf{I}}$.

\subsection{Performance Metric: Sum-Rate}
The sum-rate term for this uplink multiuser MIMO setting is given by \cite{Lozano2018}
{\setlength\abovedisplayskip{2pt}
\setlength\belowdisplayskip{2pt}
\begin{align}\label{sum_rate_function}
\mathcal{R}=\log_2\!\det\!\left(\!{\mathbf{I}}+\sum_{k=1}^{K}\frac{1}{\sigma^2}{\mathbf S}{\mathbf G}{\bm\Phi}{\mathbf H}_k{\mathbf Q}_k{\mathbf H}_k^{\mathsf{H}}
{\bm\Phi}^{\mathsf{H}}{\mathbf G}^{\mathsf{H}}{\mathbf S}^{\mathsf{H}}\!\right),
\end{align}
}where
{\setlength\abovedisplayskip{2pt}
\setlength\belowdisplayskip{2pt}
\begin{align}
{\mathbf{Q}}_k={\mathbbmss E}\left[{\mathbf x}_k{\mathbf x}_k^{\mathsf H}\right]={\mathbf{P}}_k{\mathbf{P}}_k^{\mathsf H}\in{\mathbbmss C}^{N_k\times N_k}
\end{align}
}is the transmit covariance matrix of UT $k$ subject to the power budget ${\mathsf{tr}}({\mathbf{Q}}_k)\leq p_k$. The sum-rate is achieved via minimum mean square error (MMSE) estimation along with successive interference cancellation (SIC), referred to as MMSE-SIC.

The sum-rate expression can be further simplified: by exploiting the structure of the selection matrix $\mathbf{S}$, we have
{\setlength\abovedisplayskip{2pt}
\setlength\belowdisplayskip{2pt}
\begin{align}
{\mathbf{S}}^{\mathsf{H}}{\mathbf{S}}={\mathsf{diag}}\left\{{\mathbf s}\right\}\triangleq{\bm\Delta}\in{\mathbbmss C}^{N\times N},
\end{align}
}where $\mathbf{s}=[s_1,\ldots,s_N]^{\mathsf T}$ with $s_n\in\{0,1\}$ representing the activity of antenna $n$, i.e., $s_n=1$ if antenna $n$ is selected, and $s_n=0$ otherwise. Noting ${\bm\Delta}{\bm\Delta}^{\mathsf{H}}={\bm\Delta}$ and defining $\overline{\mathbf{H}}=\left[{\mathbf G}{\bm\Phi}{\mathbf H}_1{\mathbf P}_1,\ldots,{\mathbf G}{\bm\Phi}{\mathbf H}_K{\mathbf P}_K\right]$, we can rewrite \eqref{sum_rate_function} as
{\setlength\abovedisplayskip{2pt}
\setlength\belowdisplayskip{2pt}
\begin{align}\label{Sum_Rate_Transform}
\mathcal{R}=\log_2\det\left(\mathbf{I}+{\sigma^{-2}}{\bm{\Delta}}\overline{\mathbf{H}}\overline{\mathbf{H}}^{\mathsf{H}}{\bm{\Delta}}^{\mathsf{H}}\right).
\end{align}
}In the sequel, we consider the sum-rate as the metric which quantifies the throughput of this system.

\subsection{Problem Formulation}
Our ultimate goal is to find the system design that optimizes the throughput. This means that we strive to jointly design the precoding matrices ${\mathbf P}=\{{\mathbf P}_k\}_{k=1}^{K}$\footnote{The precoding matrices can be firstly designed at the BS side and then shared with the UTs via an error-free feedback link.}, the phase-shift vector ${\bm\phi}$, and the antenna selection vector $\mathbf{s}$, such that the sum-rate term ${\mathcal R}$ is maximized. Consequently, our design problem is formulated as
{\setlength\abovedisplayskip{2pt}
\setlength\belowdisplayskip{2pt}
\begin{equation}
\label{P_1}
\begin{split}
\max_{{\mathbf s},{\mathbf P},{\bm{\phi}}}~{\mathcal R}
~{\rm{s.t.}}~&C_1:s_{n}\in\left\{0,1\right\},{\text{for}}~n\in[N],{\mathbf 1}^{\mathsf T}{\mathbf s}=T,\\
&C_2:{\mathsf{tr}}\left({\mathbf{P}}_k{\mathbf{P}}_k^{\mathsf H}\right)\leq p_k,{\text{for}}~ k\in[K],\\
&C_3:\phi_m\in{\mathcal{F}}_Q,{\text{for}}~m\in[M].
\end{split}\tag{$\mathcal P_1$}
\end{equation}
}The design problem in \eqref{P_1} in its basic form is challenging due to three main reasons. Firstly, the objective function of \eqref{P_1} is of a non-convex form. Secondly, the discrete constraints in $C_1$ and $C_3$ reduce \eqref{P_1} to an integer programming problem (which is an NP-hard problem). Finally, the presence of the phase-shifts further complicates the optimization procedure. These challenges make the optimal design computationally intractable. In the sequel, we develop an efficient framework to approximate the optimal design via a feasible computational complexity.

\section{PDD-Based Joint Design}\label{Section3}
We simplify problem \eqref{P_1} to a more tractable yet equivalent form by invoking the WMMSE framework \cite{Shi2011}. We then handle the resulting equivalent problem via the PDD that efficiently addresses nonconvex nonsmooth problems with coupling equality constraints \cite{Shi2020}.
\subsection{Reformulation of Problem \eqref{P_1}}
The principle problem \eqref{P_1} can be converted to a variational form in which the objective is replaced by a weighted mean squared error (WMSE) expression. To find this variational form, we regard the objective of \eqref{P_1} as the data rate (input-output mutual information) of the hypothetical channel described by ${\mathbf y}_{\rm{h}}={\bm{\Delta}}\overline{\mathbf{H}}{\mathbf x}_{\rm h}+{\mathbf n}_{\rm h}$. This is an hypothetical Gaussian MIMO channel whose channel matrix is ${\bm{\Delta}}\overline{\mathbf{H}}\in{\mathbbmss C}^{N\times L}$ with $L=\sum_{k=1}^{K}L_k$, and whose AWGN reads ${\mathbf n}_{\rm h}\sim{\mathcal{CN}}\left({\mathbf 0},\sigma^2{\mathbf I}\right)$. The vector ${\mathbf x}_{\rm h}\in{\mathbbmss C}^{L}$ further denotes an independent and identically distributed (i.i.d.) Gaussian transmit signal with mean zero and unit average power, i.e., ${\mathbf x}_{\rm h}\sim{\mathcal{CN}}({\mathbf 0},{\mathbf I})$. Given the output of this hypothetical Gaussian channel, i.e., ${\mathbf y}_{\rm{h}}$, let the matrix ${\mathbf U}_{\rm{h}}\in{\mathbbmss C}^{N\times L}$ be employed as the linear receiver to estimate ${\mathbf x}_{\rm h}$. In this case, the mean squared error (MSE) of the linear estimation is given by the trace of MSE matrix ${\mathbf E}_{\rm{h}}$ that is defined as
{\setlength\abovedisplayskip{2pt}
\setlength\belowdisplayskip{2pt}
\begin{subequations}
\begin{align}
{\mathbf E}_{\rm{h}}&={\mathbbmss{E}}_{{\mathbf s}_{\rm h},{\mathbf n}_{\rm h}}
\left\{\left({\mathbf U}_{\rm{h}}^{\mathsf{H}}{\mathbf y}_{\rm{h}}-{\mathbf s}_{\rm h}\right)\left({\mathbf U}_{\rm{h}}^{\mathsf{H}}{\mathbf y}_{\rm{h}}-{\mathbf s}_{\rm h}\right)^{\mathsf{H}}\right\}\\
&=\left({\mathbf U}_{\rm{h}}^{\mathsf{H}}{\bm{\Delta}}\overline{\mathbf{H}}-{\mathbf I}\right)\left({\mathbf U}_{\rm{h}}^{\mathsf{H}}{\bm{\Delta}}\overline{\mathbf{H}}-{\mathbf I}\right)^{\mathsf{H}}+\sigma^2{\mathbf U}_{\rm{h}}^{\mathsf{H}}{\mathbf U}_{\rm{h}}.\label{MSEMatrix}
\end{align}
\end{subequations}
}Invoking the WMMSE technique \cite{Shi2011}, we now introduce an auxiliary optimization variable ${\mathbf{W}}_{\rm{h}}\in{\mathbbmss{C}}^{L\times L}$ to define the following WMSE minimization problem:
{\setlength\abovedisplayskip{2pt}
\setlength\belowdisplayskip{2pt}
\begin{equation}
\label{P_2}
\begin{split}
&\min_{{\mathbf W}_{\rm{h}},{\mathbf U}_{\rm{h}},{\mathbf s},{\bm\phi},{\mathbf P}}~{\mathsf{tr}}\left\{{\mathbf W}_{\rm{h}}{\mathbf E}_{\rm{h}}\right\}-\log\det\left({\mathbf W}_{\rm{h}}\right)\\
&\quad\quad{\rm{s.t.}}~C_1,C_2,C_3.
\end{split}\tag{$\mathcal{P}_2$}
\end{equation}
}Theorem 1 in \cite{Shi2011} indicates that \eqref{P_2} is a variational form of the principle problem \eqref{P_1}. This means that the solutions $\left\{{\mathbf s},{\bm\phi},{\mathbf P}\right\}$ for both the problems are identical. It is worth mentioning that unlike the principle from in problem \eqref{P_1}, the problem \eqref{P_2} has an objective function that is marginally convex over each of the variable ${\mathbf W}_{\rm{h}}$, ${\mathbf U}_{\rm{h}}$, ${\mathbf s}$, ${\bm\phi}$, and ${\mathbf P}$.

The marginally convex form of the objective in \eqref{P_2} suggests to use the alternating optimization (AO) scheme. We however still face the issue of having discrete constraints $C_1$ and $C_3$. To tackle this issue, we define the auxiliary variables $\overline{\mathbf{s}}=[\overline{s}_1,\ldots,\overline{s}_N]^{\mathsf T}$ and ${\mathbf v}=[v_1,\ldots,v_M]^{\mathsf T}$ which satisfy the following constraints: ${\overline s}_n=s_n$, $s_n\left(1-{\overline s}_n\right)=0$, ${\mathbf v}={\bm\phi}$, and $v_m\in{\mathcal F}_Q$. We thus can equivalently find the solution of \eqref{P_2} by solving the following optimization:
{\setlength\abovedisplayskip{2pt}
\setlength\belowdisplayskip{2pt}
\begin{equation}
\label{P_3}
\begin{split}
&\min_{{\mathbf W}_{\rm{h}},{\mathbf U}_{\rm{h}},\overline{\mathbf s},{\mathbf s},{\bm\phi},{\mathbf v},{\mathbf P}}~{\mathsf{tr}}\left\{{\mathbf W}_{\rm{h}}{\mathbf E}_{\rm{h}}\right\}-\log\det\left({\mathbf W}_{\rm{h}}\right)\\
&~{\rm{s.t.}}~{\overline s}_n=s_n,s_n\left(1-{\overline s}_n\right)=0,{\text{for}}~n\in[N],{\mathbf 1}^{\mathsf T}{\mathbf s}=T,\\
&\qquad{\bm\phi}={\mathbf v},{v}_m\in{\mathcal F}_{Q},{\text{for}}~m\in[M],\\
&\qquad{\mathsf{tr}}\left({\mathbf{P}}_k{\mathbf{P}}_k^{\mathsf H}\right)\leq p_k,{\text{for}}~k\in[K].
\end{split}\tag{${\mathcal P}_3$}
\end{equation}
}It is worth mentioning that in \eqref{P_3}, variables $\mathbf s$ and $\bm\phi$ are only constrained through ${\overline s}_n=s_n$, $s_n\left(1-{\overline s}_n\right)=0$, ${\mathbf 1}^{\mathsf T}{\mathbf s}=T$, and ${\bm\phi}={\mathbf v}$. To handle these equality constraints, we resort to the PDD technique.
\subsection{The Proposed PDD-Based Algorithm}
The PDD-based algorithm is characterized by an embedded double loop structure \cite{Shi2020}. The inner loop solves the augmented Lagrangian (AL) subproblem while the outer loop updates the dual variables and the penalty parameters that correspond to constraint violation. We illustrate these concepts clearly in the sequel throughout the derivation of the PDD-based algorithm.

PDD converts problem \eqref{P_3} into its AL by moving the equality constraints as a penalty term to the objective function. The AL subproblem corresponding to \eqref{P_3} is given by
{\setlength\abovedisplayskip{2pt}
\setlength\belowdisplayskip{2pt}
\begin{equation}
\label{P_4}
\begin{split}
&\min_{{\mathbf W}_{\rm{h}},{\mathbf U}_{\rm{h}},\overline{\mathbf s},{\mathbf s},{\bm\phi},{\mathbf v},{\mathbf P}}~{\mathsf{tr}}\left\{{\mathbf W}_{\rm{h}}{\mathbf E}_{\rm{h}}\right\}-\log\det\left({\mathbf W}_{\rm{h}}\right)+f_{\rho}\\
&~{\rm{s.t.}}~{v}_m\in{\mathcal F}_{Q},{\text{for}}~m\in[M],\\
&~\quad~~{\mathsf{tr}}\left({\mathbf{P}}_k{\mathbf{P}}_k^{\mathsf H}\right)\leq p_k,{\text{for}}~k\in[K],
\end{split}\tag{${\mathcal P}_4$}
\end{equation}
}where $\rho>0$ is the penalty parameter penalizing the violation of the equality constraints, and $f_\rho$ is given by
{\setlength\abovedisplayskip{2pt}
\setlength\belowdisplayskip{2pt}
\begin{align}
&f_{\rho}=\frac{1}{2\rho}\left[\left({\mathbf 1}^{\mathsf T}{\mathbf s}-T+\rho\xi\right)^2
+\sum_{n=1}^{N}\left[\left(s_n-{\overline s}_n+\rho\mu_n\right)^2\right.\right.\nonumber\\
&\left.\left.+\left(s_n\left(1-{\overline s}_n\right)+\rho\lambda_n\right)^2\right]+\sum_{m=1}^{M}|\phi_m-v_m+\rho\tau_m|^2\right],
\end{align}
}with $\xi$, ${\bm\mu}=[\mu_1,\ldots,\mu_N]^{\mathsf T}$, ${\bm\lambda}=[\lambda_1,\ldots,\lambda_N]^{\mathsf T}$, and ${\bm\tau}=[\tau_1,\ldots,\tau_M]^{\mathsf T}$ denting the dual variables associated with the equality constraints in ${\mathbf 1}^{\mathsf T}{\mathbf s}=T$, ${\overline s}_n=s_n$, $s_n\left(1-{\overline s}_n\right)=0$, and ${\bm\phi}={\mathbf v}$, respectively. It is observed that as $\rho\rightarrow0$, the penalty term is forced to zero, i.e., equality constraints are enforced. It is shown in \cite{Shi2020} that updating the primal and dual variables, as well as the penalty factor in an alternating manner, PDD converges to a stationary-point solution.

Starting with the penalized form \eqref{P_4}, we now discuss the underlying problems in the inner and outer loops.
The solution to the AL problem (inner loop) is iteratively approximated via the block coordinate descent (BCD) method. To this end, we minimize the objective function of \eqref{P_4} by sequentially updating ${\mathbf W}_{\rm{h}}$, ${\mathbf U}_{\rm{h}}$, ${\mathbf s}$, $\overline{\mathbf s}$, ${\bm\phi}$, $\mathbf v$, and $\mathbf P$ in the inner loop through the following marginal optimizations.
\subsubsection{Optimizing the Weights}
We optimize ${\mathbf W}_{\rm{h}}$ while treating the remaining variables as constants, i.e., we find
{\setlength\abovedisplayskip{2pt}
\setlength\belowdisplayskip{2pt}
\begin{align}
{\mathbf W}_{\rm{h}}^{\star}=\argmin\nolimits_{{\mathbf W}_{\rm{h}}}{\mathsf{tr}}\left\{{\mathbf W}_{\rm{h}}{\mathbf E}_{\rm{h}}\right\}-\log\det\left({\mathbf W}_{\rm{h}}\right).
\end{align}
}Using its first-order optimality condition, we get ${\mathbf W}_{\rm{h}}^{\star}={\mathbf E}_{\rm{h}}^{-1}$.
\subsubsection{Optimizing the Hypothetical Receiver}
The marginally optimal ${\mathbf U}_{\rm{h}}$ is given by solving a linear MSE minimization problem whose solution is
{\setlength\abovedisplayskip{2pt}
\setlength\belowdisplayskip{2pt}
\begin{align}
{\mathbf U}_{\rm{h}}^{\star}=(\sigma^2{\mathbf{I}}+{\bm{\Delta}}\overline{\mathbf{H}}{\overline{\mathbf{H}}}^{\mathsf{H}}{\bm\Delta}^{\mathsf{H}})^{-1}{\bm{\Delta}}\overline{\mathbf{H}}.
\end{align}
}\subsubsection{Optimizing $\overline{\mathbf s}$}
The marginal optimization for $\{\overline{s}_n\}_{n=1}^{N}$ decouples into $N$ scalar optimization problems with solution to the $n$th problem being
{\setlength\abovedisplayskip{2pt}
\setlength\belowdisplayskip{2pt}
\begin{equation}\label{Problem_Auxiliary}
\overline{s}_n^{\star}=\argmin\nolimits_{\overline{s}_n\in{\mathbbmss R}}\left(\overline{a}_n\overline{s}_n^2-2\overline{b}_n\overline{s}_n\right).
\end{equation}
}Here, $\overline{a}_n=1+s_n^2$ and $\overline{b}_n=s_n+\rho\lambda_n+s_n^2+s_n\rho\mu_n$. The optimization in \eqref{Problem_Auxiliary} is a quadratic scalar problem whose solution is given by ${\overline{s}}_n^{\star}={\overline{b}_n}/{\overline{a}_n}$.
\subsubsection{Optimizing $\mathbf s$}
The marginal problem for $\mathbf s$ is given by
{\setlength\abovedisplayskip{2pt}
\setlength\belowdisplayskip{2pt}
\begin{equation}\label{Beam_Selection_Matrix_Opt}
{\mathbf s}^{\star}=\argmin\nolimits_{{\mathbf s}\in{\mathbbmss R}^{N}}\left({\mathbf s}^{\mathsf T}{\mathbf Q}{\mathbf s}-{\mathbf s}^{\mathsf T}{\mathbf g}\right),
\end{equation}
}where
{\setlength\abovedisplayskip{2pt}
\setlength\belowdisplayskip{2pt}
\begin{align}
{\mathbf Q}&\triangleq
\Re\left\{\left({\mathbf U}_{\rm{h}}{\mathbf W}_{\rm{h}}{\mathbf U}_{\rm{h}}^{\mathsf{H}}\right)\odot(\overline{\mathbf{H}}\overline{\mathbf{H}}^{\mathsf{H}})\right\}+{2^{-1}\rho^{-1}}\left({\mathbf 1}{\mathbf 1}^{\mathsf{T}}+{\mathbf I}\right)\nonumber\\
&\quad+{2^{-1}\rho^{-1}}{{\mathsf{diag}}\left\{\left(1-{\overline s}_1\right)^2,\cdots,\left(1-{\overline s}_N\right)^2\right\}},\\
{\mathbf g}&\triangleq2\Re\!\left\{{\mathbf q}\right\}\!-\!\frac{1}{\rho}\left(\left(\rho\xi\!-\!T\right){\mathbf 1}\!+\!\left(\rho{\bm\lambda}\!-\!\overline{\mathbf s}\right)\!+\!\rho\left({\mathbf{1}}\!-\!\overline{\mathbf s}\right)\!\odot\!{\bm\mu}\right),\\
{\mathbf q}&\triangleq\left[[{\mathbf U}_{\rm{h}}{\mathbf W}_{\rm{h}}\overline{\mathbf{H}}^{\mathsf{H}}]_{1,1},\cdots,[{\mathbf U}_{\rm{h}}{\mathbf W}_{\rm{h}}\overline{\mathbf{H}}^{\mathsf{H}}]_{N,N}\right]^{\mathsf{T}}.
\end{align}
}Using the Schur product theorem, we can conclude that $({\mathbf U}_{\rm{h}}{\mathbf W}_{\rm{h}}{\mathbf U}_{\rm{h}}^{\mathsf{H}})\odot(\overline{\mathbf{H}}\overline{\mathbf{H}}^{\mathsf{H}})\succeq{\mathbf 0}$ and thus
${\mathbf Q}\succ{\mathbf 0}$, which means that the problem \eqref{Beam_Selection_Matrix_Opt} is a standard convex problem. Using its first-order optimality condition, we obtain ${{\mathbf s}}^{\star}=\frac{1}{2}{\mathbf Q}^{-1}{\mathbf g}$.
\subsubsection{Optimizing the Phase-Shifts}
The marginal problem for ${\bm\phi}$ is given by
{\setlength\abovedisplayskip{2pt}
\setlength\belowdisplayskip{2pt}
\begin{align}\label{P_6}
{\bm\phi}^{\star}=\argmin\nolimits_{{\bm\phi}\in{\mathbbmss C}^{M}}~({\bm\phi}^{\mathsf{H}}\mathbf{B}{\bm\phi}-2\Re\{{\bm\phi}^{\mathsf{H}}{\mathbf{b}}\}).
\end{align}
}The terms appearing in \eqref{P_6} are defined as follows:
\vspace{-5pt}
\begin{itemize}
  \item ${\mathbf{b}}=\mathsf{diag}\{{\mathbf{G}}^{\mathsf{H}}{\bm\Delta}^{\mathsf{H}}{\mathbf{U}}_{\rm{h}}{\mathbf{W}}_{\rm{h}}{\mathbf{K}}^{\mathsf{H}}\}+\frac{1}{2\rho}({\mathbf v}-\rho{\bm\tau})$.
  \item ${\mathbf{K}}=\left[{\mathbf H}_1{\mathbf P}_1,\ldots,{\mathbf H}_K{\mathbf P}_K\right]$.
  \item ${\mathbf{B}}={\mathbf{B}}_1\odot{\mathbf{B}}_2^{\mathsf{T}}+\frac{1}{2\rho}{\mathbf I}$.
  \item ${\mathbf{B}}_1={\mathbf{G}}^{\mathsf{H}}{\bm\Delta}^{\mathsf{H}}{\mathbf{U}}_{\rm{h}}{\mathbf{W}}_{\rm{h}}{\mathbf{U}}_{\rm{h}}^{\mathsf{H}}
{\bm\Delta}{\mathbf{G}}\succeq{\mathbf 0}$ and ${\mathbf{B}}_2={\mathbf{K}}{\mathbf{K}}^{\mathsf{H}}\succeq{\mathbf 0}$.
\end{itemize}
\vspace{-5pt}
Using the Schur product theorem, we can conclude that ${\mathbf{B}}_1\odot{\mathbf{B}}_2^{\mathsf{T}}\succeq{\mathbf 0}$, and thus
${\mathbf B}\succ{\mathbf 0}$. Therefore, problem \eqref{P_6} is an unconstrained convex quadratic optimization problem. Using its first-order optimality condition, we find the solution as
{\setlength\abovedisplayskip{2pt}
\setlength\belowdisplayskip{2pt}
\begin{align}
{\bm\phi}^{\star}={\mathbf{B}}^{-1}{\mathbf{b}}.
\end{align}
}\subsubsection{Optimizing $\mathbf v$}
The marginal optimization for this auxiliary variable $\mathbf v$ decouples into $M$ parallel scalar optimizations with the $m$th subproblem being
{\setlength\abovedisplayskip{2pt}
\setlength\belowdisplayskip{2pt}
\begin{align}
v_{m}^{\star}&=\argmin\nolimits_{v_m\in{\mathcal{F}}_Q}|\phi_m-v_m+\rho\tau_m|.
\end{align}
}Using the fact that $|v_m|^2=1$, we have
{\setlength\abovedisplayskip{2pt}
\setlength\belowdisplayskip{2pt}
\begin{align}
v_{m}^{\star}&=\argmax\nolimits_{v_m\in{\mathcal{F}}_Q}\Re\{v_m(\phi_m+\rho\tau_m)^{*}\}={\rm e}^{{\rm j}{\vartheta}_m^{\star}},
\end{align}
}where ${\vartheta}_m^{\star}$ is obtained by mapping $\angle({\phi_m+\rho\tau_m})$ to the nearest discrete phase in the set of discrete phases. With infinite-precision representation of phase-shifts, i.e., $Q\rightarrow\infty$, we have $v_{m}^{\star}={\rm{e}}^{{\rm j}\angle({\phi_m+\rho\tau_m})}$.
\subsubsection{Optimizing the Precoders}
The marginal optimization in terms of ${\mathbf P}$ reduces to $K$ subproblems with the $k$th one being
{\setlength\abovedisplayskip{2pt}
\setlength\belowdisplayskip{2pt}
\begin{align}
{\mathbf P}_k^{\star}\!=\!\argmin\nolimits_{{\mathsf{tr}}({\mathbf{P}}_k{\mathbf{P}}_k^{\mathsf H})\leq p_k}({\mathsf{tr}}({\mathbf P}_k^{\mathsf H}{\mathbf C}_k{\mathbf P}_k)\!-\!2\Re\{{\mathsf{tr}}({\mathbf P}_k^{\mathsf H}{\mathbf D}_k)\}),\nonumber
\end{align}
}where ${\mathbf C}_k={\mathbf O}_k{\mathbf{W}}_{\rm h}{\mathbf O}_k^{\mathsf H}$, ${\mathbf O}_k={\mathbf H}_k^{\mathsf H}{\bm\Phi}^{\mathsf H}{\mathbf G}^{\mathsf H}{\bm\Delta}^{\mathsf H}{\mathbf U}_{\rm h}$, ${\mathbf D}_k={\mathbf O}_k{\mathbf{W}}_{{\rm h},k}^{\mathsf H}$, and ${\mathbf{W}}_{{\rm h},k}$ denotes the submatrix of ${\mathbf{W}}_{{\rm h}}$ obtained by extracting the elements of the rows with indices from $(1+\sum_{k'=1}^{k-1}L_{k'})$ to $\sum_{k'=1}^{k}L_{k'}$. This is a standard convex quadratic optimization subproblem whose solution is given by
{\setlength\abovedisplayskip{2pt}
\setlength\belowdisplayskip{2pt}
\begin{align}
{\mathbf P}_k^{\star}=({\mathbf{C}}_k+\lambda{\mathbf I})^{-1}{\mathbf{D}}_k.
\end{align}
}The regularizer $\lambda$ is chosen, such that the complementarity slackness condition, i.e., $\lambda({\mathsf{tr}}({\mathbf{P}}_k{\mathbf{P}}_k^{\mathsf H})- p_k)=0$, is satisfied. If ${\mathsf{tr}}({\mathbf{D}}_k^{\mathsf H}({\mathbf{C}}_k+\lambda{\mathbf I})^{-2}{\mathbf{D}}_k)=p_k$; then, $\lambda=0$. Otherwise, we can obtain the solution of $\lambda$ from the following identity:
{\setlength\abovedisplayskip{2pt}
\setlength\belowdisplayskip{2pt}
\begin{align}
{\mathsf{tr}}({\mathbf{P}}_k{\mathbf{P}}_k^{\mathsf H})={\mathsf{tr}}({\mathbf{D}}_k^{\mathsf H}({\mathbf{C}}_k+\lambda{\mathbf I})^{-2}{\mathbf{D}}_k)=p_k.
\end{align}
}It follows that
{\setlength\abovedisplayskip{2pt}
\setlength\belowdisplayskip{2pt}
\begin{align}
\sum\nolimits_{n_k=1}^{N_k}\frac{|[{\mathbf U}_{{\mathbf C}_k}{\mathbf D}_k{\mathbf D}_k^{\mathsf H}{\mathbf U}_{{\mathbf C}_k}^{\mathsf H}]_{n_k,n_k}|^2}{([{\bm\Lambda}_{{\mathbf C}_k}]_{n_k,n_k}+\lambda)^{2}}=p_k,\label{Power_Bisection}
\end{align}
}where ${\mathbf C}_k={\mathbf U}_{{\mathbf C}_k}^{\mathsf H}{\bm\Lambda}_{{\mathbf C}_k}{\mathbf U}_{{\mathbf C}_k}$ is the eigen-decomposition of ${\mathbf C}_k$. Since $[{\bm\Lambda}_{{\mathbf C}_k}]_{n_k,n_k}\geq0$ for $n_k\in[N_k]$, the left-hand side of \eqref{Power_Bisection} is a monotonously decreasing function of $\lambda\geq0$. Consequently, we can find $\lambda$ by solving equation \eqref{Power_Bisection} using the bisection-based search method.

We next consider the outer loop of the PDD-based algorithm. In this loop, the dual variables $\{\xi,{\bm\mu},{\bm\lambda},{\bm\tau}\}$ and the penalty factor $\rho$ are updated. These variables are updated by optimizing the constraint violation quantity that is defined as
{\setlength\abovedisplayskip{2pt}
\setlength\belowdisplayskip{2pt}
\begin{align}\label{Constraint_Vio}
h\triangleq\max_{\forall n,m}\{
|\phi_m\!-\!v_m|,|{\mathbf 1}\!^{\mathsf T}{\mathbf s}\!-\!T|,\left|{\overline s}_n\!-\!s_n\right|,\left|s_n\!\left(1\!-\!{\overline s}_n\right)\right|\}.
\end{align}
}The optimization of this quantity is discussed in great detail in \cite{Shi2020} and is skipped here, due to lack of space. The outcome of this optimization yield the update scheme as follows: if $h<\mu$, we update the dual variables via the recursive equalities
{\setlength\abovedisplayskip{2pt}
\setlength\belowdisplayskip{2pt}
\begin{subequations}\label{Dual_Variable_Update}
\begin{align}
\xi^{(t+1)}&=\xi^{(t)}+{\rho}^{-1}\left({\mathbf 1}^{\mathsf T}{\mathbf s}-T\right),\\
\mu_n^{(t+1)}&=\mu_n^{(t)}+{\rho}^{-1}\left({\overline s}_n-s_n\right),\\
\lambda_n^{(t+1)}&=\lambda_n^{(t)}+{\rho}^{-1}s_n\left(1-{\overline s}_n\right),\\
\tau_m^{(t+1)}&=\tau_m^{(t)}+{\rho}^{-1}(v_m-\phi_m),
\end{align}
\end{subequations}
}where $t$ denotes the outer iteration index. We refer the reader to \cite[Table \uppercase\expandafter{\romannumeral1}, Line 4]{Shi2020} for further details. The derived PDD-based algorithm is summarized in Algorithm \ref{Algorithm2}.

\subsection{Convergence and Complexity Analyses}
As shown in \cite{Shi2020}, Algorithm \ref{Algorithm2} is guaranteed to converge to a stationary solution of problem \eqref{P_1}. The computational complexity of the proposed algorithm can be further characterized in terms of problem dimensions. To this end, let $I_{\rm{out}}$ and $I_{\rm{in}}$ denote the numbers of iterations in the outer loop and the inner loop, respectively. The per-iteration complexity in the outer loop is mainly originated from the BCD updates in the inner loop. The computational complexity of one inner-loop iteration is further composed of the complexity of updating variables $\left\{{\mathbf W}_{\rm{h}},{\mathbf U}_{\rm{h}},\bar{\mathbf{s}},{\mathbf{s}},{\bm\phi},{\mathbf v},{\mathbf P}\right\}$. It is readily shown that the complexity of marginal optimizations with respect to ${\mathbf W}_{\rm{h}}$, ${\mathbf U}_{\rm{h}}$, $\bar{\mathbf s}$, ${\mathbf s}$, ${\bm\phi}$, $\mathbf v$, and $\mathbf P$ scale with ${\mathcal O}\left(L^3\right)$, ${\mathcal O}\left(N^3\right)$, ${\mathcal O}\left(N\right)$, ${\mathcal O}\left(N^3\right)$, ${\mathcal O}\left(M^3\right)$, ${\mathcal O}\left(M\right)$, and ${\mathcal O}\left(\sum_{k=1}^{K}N_k^3\right)$, respectively. Hence, the overall complexity of Algorithm \ref{Algorithm2} scales with ${\mathcal O}\left(I_{\text{out}}I_{\text{in}}\left(2N^3+M^3+L^3+\sum_{k=1}^{K}N_k^3\right)\right)$, which is of a polynomial order.

\begin{algorithm}[htbp]
  \caption{PDD-based algorithm for solving problem \eqref{P_1}}
  \label{Algorithm2}
  \begin{algorithmic}[1]
    \STATE Initialize primary variables $\left\{{\mathbf P},{\mathbf v},{\bm\phi},{\mathbf{s}},\overline{\mathbf{s}},{\mathbf W}_{\rm{h}},{\mathbf U}_{\rm{h}}\right\}$, dual variables $\left\{\xi,{\bm\lambda},{\bm\mu},{\bm\tau}\right\}$, iteration index $t=0$, threshold $\mu$, penalty factor $\rho>0$, and scaling factor $\chi\in(0,1)$
    \REPEAT
    \REPEAT
      \STATE Update $\left\{{\mathbf W}_{\rm{h}},{\mathbf U}_{\rm{h}},\overline{\mathbf{s}},{\mathbf{s}},{\bm\phi},{\mathbf v},{\mathbf P}\right\}$ by the BCD method
    \UNTIL{convergence}
    \STATE Calculate the constraint violation $h$ by \eqref{Constraint_Vio}
    \IF{$h < \mu$}
	\STATE Update the dual variable by \eqref{Dual_Variable_Update}
	\ELSE
	\STATE Set $\rho = \chi\rho$
	\ENDIF
    \STATE Set $\mu=\chi h$ and $t=t+1$
    \UNTIL{convergence}
  \end{algorithmic}
\end{algorithm}

\section{Low-Complexity Sequential Optimization}
The PDD-based algorithm proposed in Section \ref{Section3} optimizes the antenna selection vector $\mathbf s$, phase-shift vector $\bm\phi$, and transmit precoders $\{{\mathbf P}_k\}_{k=1}^{K}$ alternately. Although this leads to a feasible complexity, it can lead to a computational burdensome in many applications. To this end, we develop an alternative scheme based on sequential optimization approach in this section. In this respect, we first design the phase-shifts using an element-wise BCD (EWBCD) method; then, we select the antennas using a greedy search (GS). We finally obtain the optimal transmit precoders with the aid of iterative water-filling method. These three steps are illustrated in the sequel.
\subsubsection{Phase-Shifts Design}
At the first step of the SO-based method, we design the phase-shifts by maximizing the effective channel gain given by
{\setlength\abovedisplayskip{2pt}
\setlength\belowdisplayskip{2pt}
\begin{align}
\mathsf{tr}({\overline{\mathbf{H}}}{\overline{\mathbf{H}}}^{\mathsf H})=\sum\nolimits_{k=1}^{K}\mathsf{tr}({\mathbf G}{\bm\Phi}{\mathbf H}_k{\mathbf Q}_k{\mathbf H}_k^{\mathsf H}{\bm\Phi}^{\mathsf H}{\mathbf G}^{\mathsf H}).
\end{align}
}Assuming ${\mathbf Q}_k={\frac{p_k}{N_k}}{\mathbf I}$ for $k\in[K]$, the phase-shifts design problem is formulated as follows:
{\setlength\abovedisplayskip{2pt}
\setlength\belowdisplayskip{2pt}
\begin{align}\label{SO_1}
{\bm\phi}^{\star}=\argmax({\mathsf{tr}}({\bm\Phi}^{\mathsf H}{\mathbf G}^{\mathsf H}{\mathbf G}{\bm\Phi}\widetilde{\mathbf H})={\bm\phi}^{\mathsf H}\hat{\mathbf H}{\bm\phi}),~{\rm{s.t.}}~C_3,
\end{align}
}where $\widetilde{\mathbf H}=\sum\nolimits_{k=1}^{K}\frac{p_k}{N_k}{\mathbf H}_k{\mathbf H}_k^{\mathsf H}$ and $\hat{\mathbf H}=({\mathbf G}^{\mathsf H}{\mathbf G})\odot{\widetilde{\mathbf H}}^{\mathsf T}$. To handle the discretely constrained $\bm\phi$, we propose to optimize the elements of $\bm\phi$ one by one, resulting in an EWBCD method comprised of $M$ steps. At the $m$th step, we optimize $\phi_m$ while treating $\phi_{m'}$ for $m'\neq m$ as constants. This leads to
{\setlength\abovedisplayskip{2pt}
\setlength\belowdisplayskip{2pt}
\begin{align}\label{SO_1_Opt}
{\phi}_m^{\star}=\argmax\nolimits_{\phi_m\in{\mathcal F}_Q}{2\Re\{\phi_m^{*}\varphi_m\}},
\end{align}
}where $\varphi_m=\sum\nolimits_{m'\neq m}[\hat{\mathbf H}]_{m',m}\phi_{m'}$. The solution of this problem is given by ${\phi}_m^{\star}={\rm e}^{{\rm j}\delta_m^{\star}}$, where $\delta_m^{\star}$ is obtained by mapping $\angle{\varphi_m}$ to the nearest discrete phase.
\subsubsection{Antenna Selection}
In the next step, we optimize the antenna selection vector $\mathbf s$ for the updated phase-shifts. The resulting problem of antenna selection is given by
{\setlength\abovedisplayskip{2pt}
\setlength\belowdisplayskip{2pt}
\begin{align}\label{SO_2}
{\mathbf s}^{\star}=\argmax\log_2\det(\mathbf{I}+{\sigma^{-2}}{\bm{\Delta}}{\mathbf X}{\bm{\Delta}}^{\mathsf{H}}),~{\rm{s.t.}}~C_1,
\end{align}
}where ${\mathbf X}={\mathbf G}{\bm\Phi}\widetilde{\mathbf H}{\bm\Phi}^{\mathsf H}{\mathbf G}^{\mathsf H}$. To tackle the discrete constraint, we exploit the GS method. More details about this method can be found in \cite{Gershman2004,Ali2018} and are omitted here due to page limitations.
\subsubsection{Transmit Precoding Design}
After obtaining the phase-shifts and antenna selection vector, we find the optimal transmit precoding matrices by solving
{\setlength\abovedisplayskip{2pt}
\setlength\belowdisplayskip{2pt}
\begin{align}\label{SO_3}
\max_{\mathbf P}~\log_2\det(\mathbf{I}+\sum\nolimits_{k=1}^{K}{\mathbf G}_{{\bm\phi},k}{\mathbf{P}}_k{\mathbf{P}}_k^{\mathsf H}{\mathbf G}_{{\bm\phi},k}^{\mathsf{H}}),~{\rm{s.t.}}~C_2,
\end{align}
}where ${\mathbf G}_{{\bm\phi},k}=\sigma^{-1}{\bm\Delta}{\mathbf G}{\bm\Phi}{\mathbf H}_k$. The optimization in \eqref{SO_3} can be optimally solved by the iterative water-filling method \cite{Yu2004}.

Based on \cite{Gershman2004} and \cite{Yu2004}, we can show that the complexity of sequential optimizations with respect to ${\bm\phi}$, ${\mathbf s}$, and $\mathbf P$ scale with ${\mathcal O}\left(M\right)$, ${\mathcal O}\left(NTM\right)$, and ${\mathcal O}\left(KT^3\right)$, respectively. Hence, the complexity of the SO-based method scales with ${\mathcal O}(NTM+KT^3)$, which is lower than the PDD-based method.

\begin{figure}[!t]
    \centering
    \subfigbottomskip=0pt
	\subfigcapskip=-5pt
\setlength{\abovecaptionskip}{0pt}
    \subfigure[Convergence of sum rate.]
    {
        \includegraphics[height=0.18\textwidth]{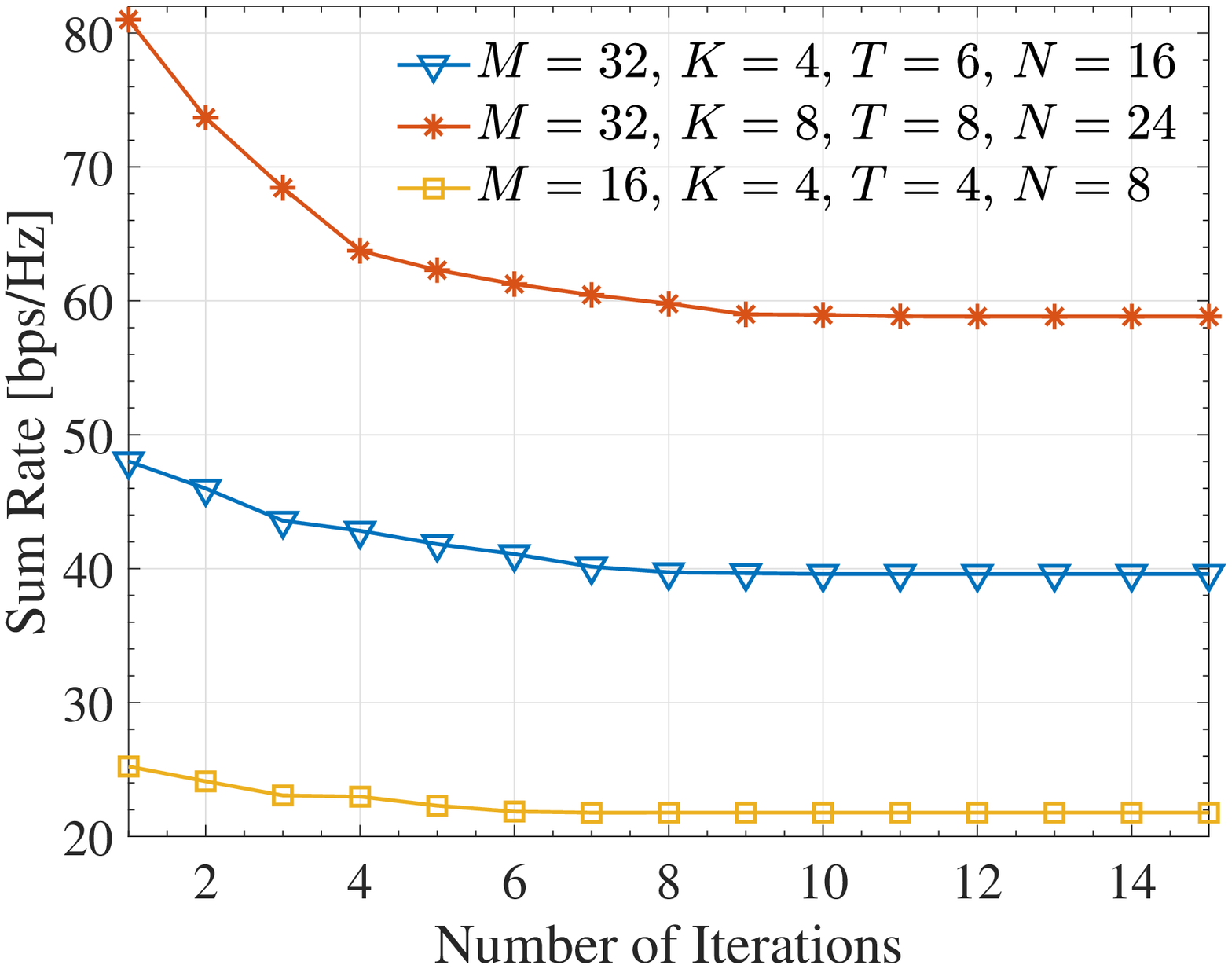}
	   \label{fig1a}	
    }\hspace{-9pt}
   \subfigure[Constraint violation.]
    {
        \includegraphics[height=0.18\textwidth]{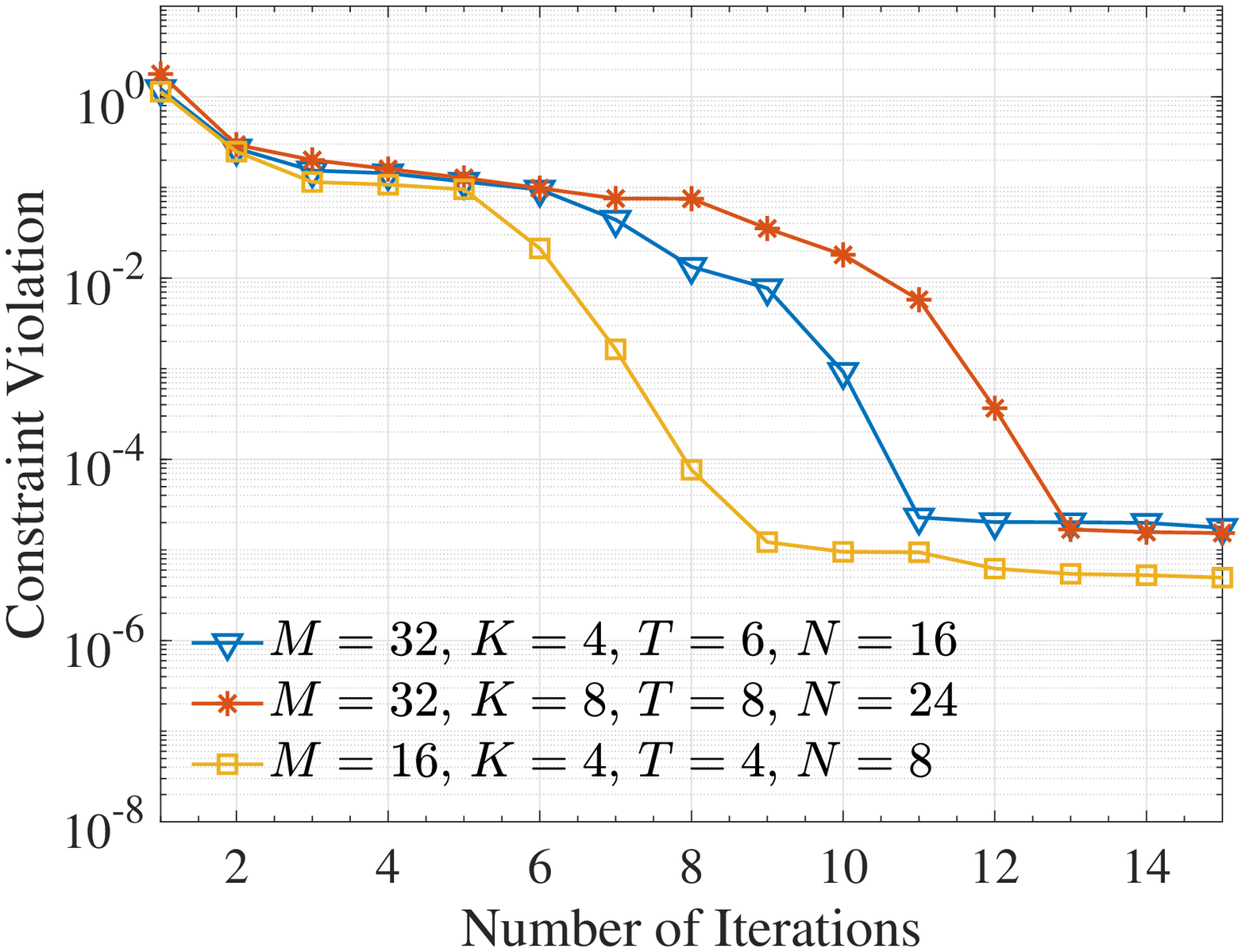}
	   \label{fig1b}	
    }
\caption{Average convergence performances for $Q=4$ and $\log{p_k}=-5$ dBm ($\forall k$).}
    \label{figure1}
    \vspace{-20pt}
\end{figure}

\begin{figure*}[!t]
    \centering
    \subfigbottomskip=0pt
	\subfigcapskip=-5pt
\setlength{\abovecaptionskip}{0pt}
   \subfigure[Sum rate vs. the power budget.]
    {
        \includegraphics[height=0.23\textwidth]{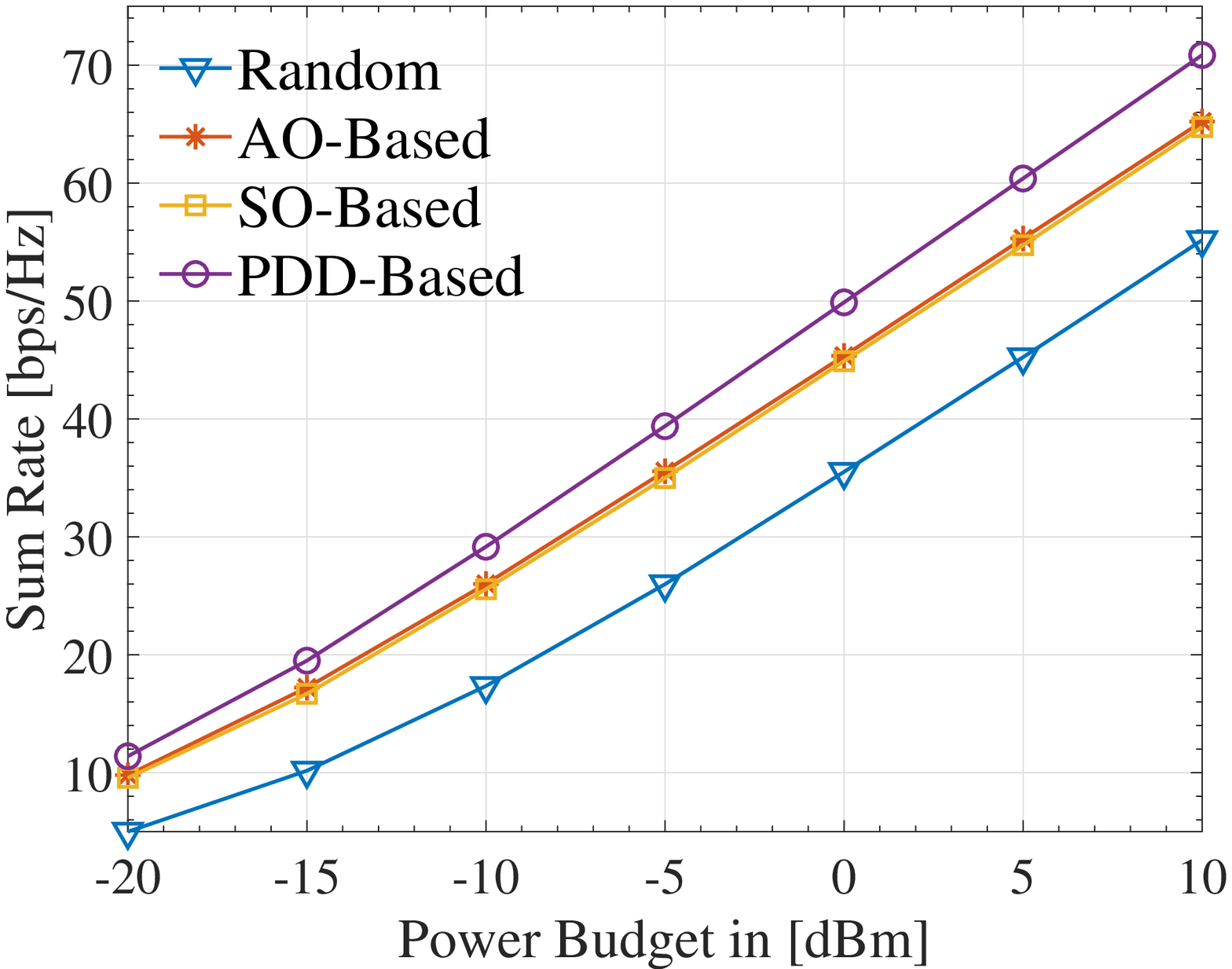}
	   \label{fig2a}	
    }\hspace{-5pt}
    \subfigure[Sum rate vs. the number of quantization bits.]
    {
        \includegraphics[height=0.23\textwidth]{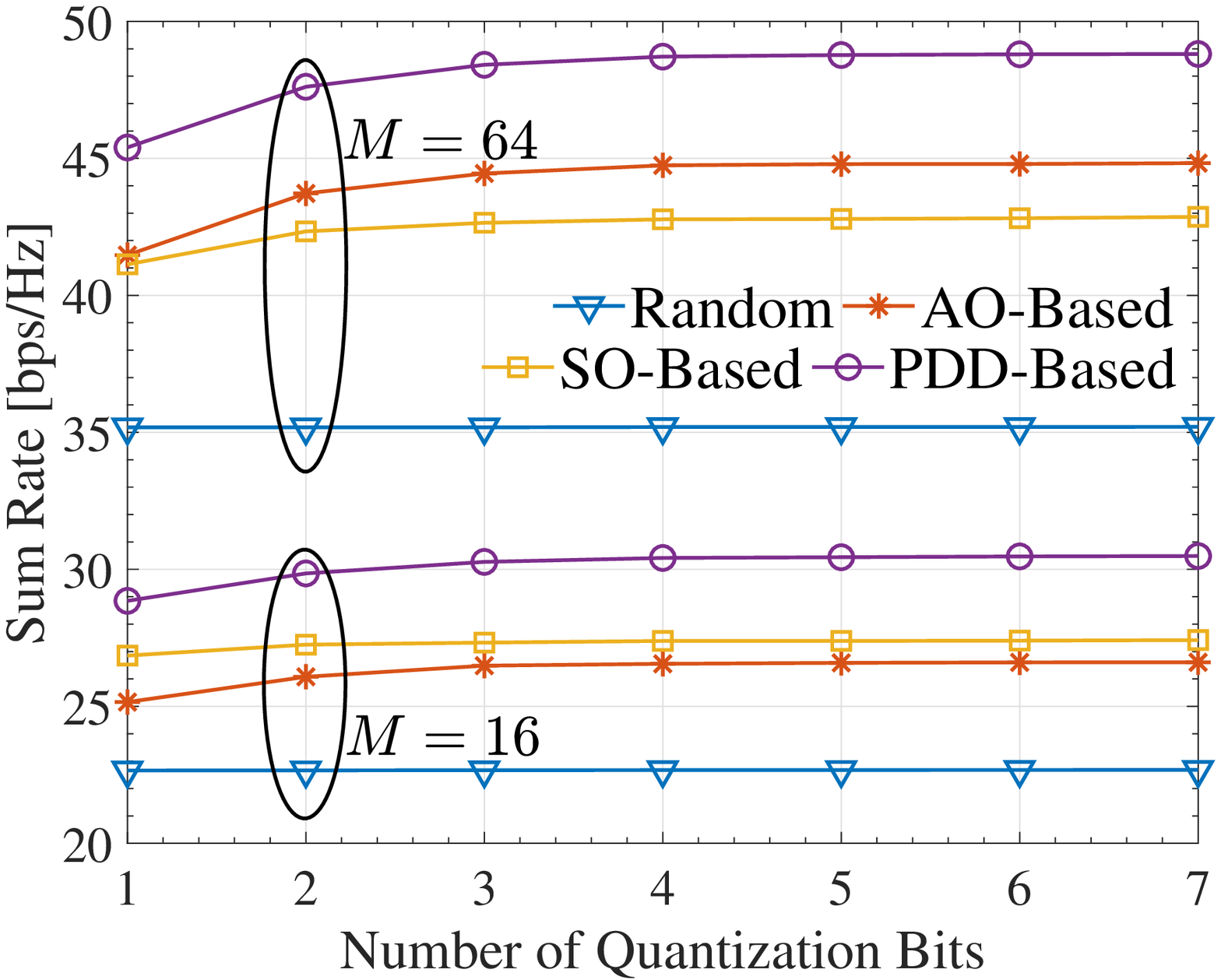}
	   \label{fig2b}	
    }\hspace{-5pt}
    \subfigure[Sum rate vs. the number of RF chains.]
    {
        \includegraphics[height=0.23\textwidth]{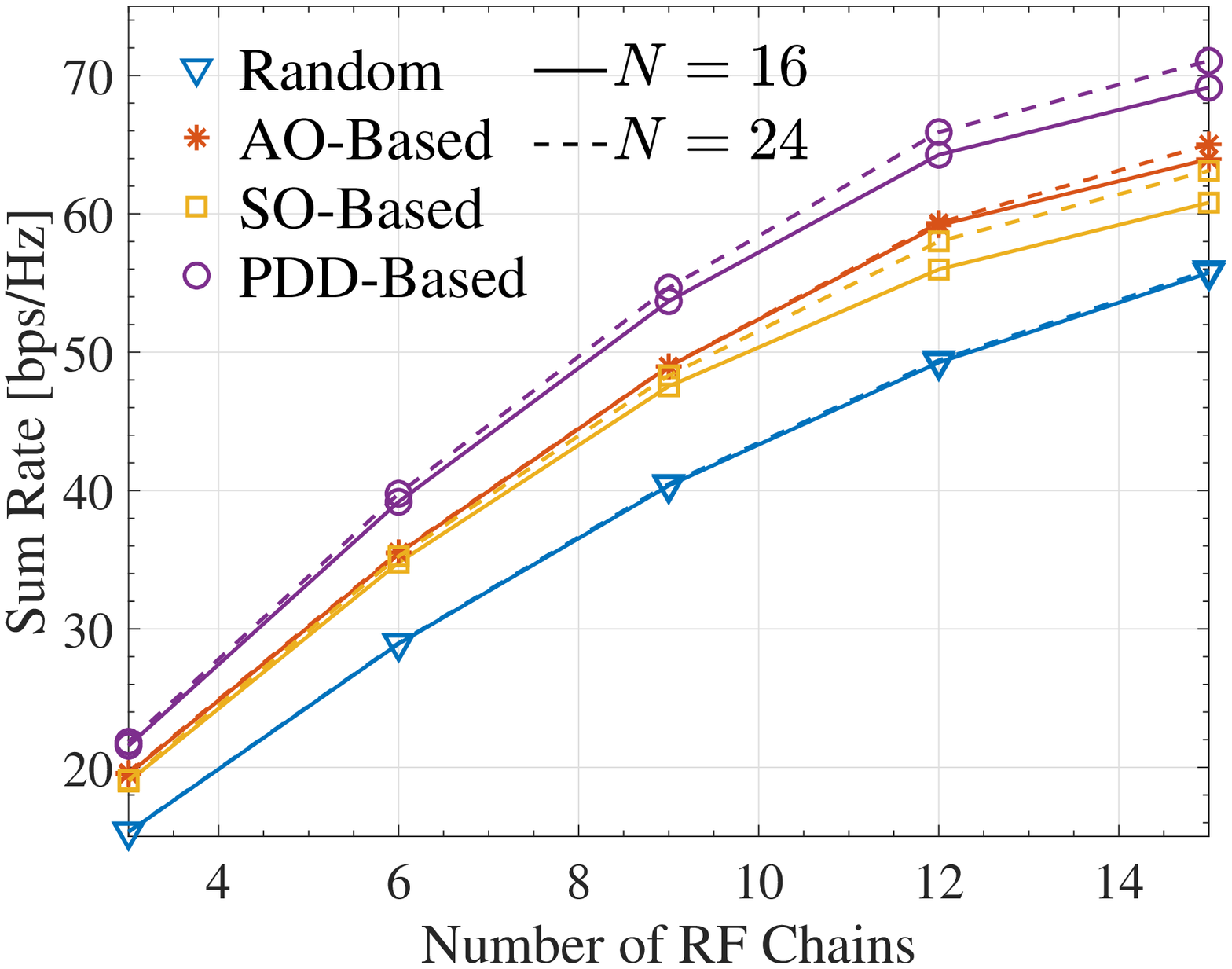}
	   \label{fig2c}	
    }
   \caption{Sum-rate performances of the proposed algorithms. The simulation parameters are given as follows: (a) $N=16$, $T=6$, $M=32$, and $Q=4$; (b) $N=16$, $T=6$, and $\log{p}=-5$ dBm; (c) $M=32$, $Q=4$, and $\log{p}=-5$ dBm.}
    \label{figure2}
    \vspace{-20pt}
\end{figure*}

\section{Numerical Results}
In this section, we verify the effectiveness of the proposed algorithms through numerical experiments. For simulations, the following parameters are used unless specified otherwise later: $N_k=L_k=3$, $p_k=p$, for $k\in[K]$, $\log{\sigma^2}=-120$ dBm, $K=4$, $\mu=1$, $\rho=1$, and $\chi=0.1$. We further generate the channel realizations as follows. Regarding the large scale fading, we assume that all the composite UT-RIS-BS channels, i.e., ${\mathbf G}{\mathbf H}_k$ exhibit the same path loss $-120$ dB for illustration. Meanwhile, for the small scale fading of the UT-to-RIS and RIS-to-BS channels, we consider the standard Rayleigh fading model. All the optimization variables are randomly initialized, and all simulation curves are averaged over $500$ independent channel realizations.

In {\figurename} {\ref{figure1}}, we first study the convergence behavior of the proposed PDD-based method. As can be seen from {\figurename} {\ref{fig1a}}, the sum-rate rapidly converges to a stationary value. {\figurename} {\ref{fig1b}} shows the constraint violation in terms of the number of outer iterations. We observe that the constraint violation reduces to a threshold $10^{-4}$ in less than $15$ outer iterations, implying that the solution has essentially met the equality constraints for problem \eqref{P_1}.

We next consider the following benchmark schemes for performance comparison: 1) Random scheme, in which $\mathbf s$ and $\bm\phi$ are randomly set, and ${\mathbf Q}_k={\frac{p_k}{N_k}}{\mathbf I}$ for $k\in[K]$. 2) AO-based scheme, in which $\mathbf s$, $\bm\phi$, and $\mathbf P$ are updated alternately until convergence. Here, we design $\mathbf s$, $\bm\phi$, and $\mathbf P$ using the GS-based, EWBCD-based, and iterative water-filling methods, respectively. This scheme is a generalization of the methods in \cite{Wang2022,Abdullah2022,Rezaei2022,He2022,Xu2022}, whose complexity scales with ${\mathcal O}(I_{\rm{AO}}(M^3+NTM+KT^3))$ with $I_{\rm{AO}}$ denotes the number of AO iterations. It is worth noting that the complexity of the AO-based method is between the PDD-based and the SO-based methods.

{\figurename} {\ref{figure2}} compares the sum-rate performances achieved by the proposed methods and the benchmark methods. In {\figurename} {\ref{fig2a}}, the sum-rate is plotted against the transmit power budget $p$ for different optimization schemes assuming $M=32$. From the figure, it is seen that the proposed PDD-based method significantly outperforms the AO-based one. The SO-based method further achieves almost the same performance as the AO-based scheme and achieves a significant performance gain as compared with the random scheme. As mentioned, the SO-based method involves less complexity than the AO-based scheme. The simulation results hence imply that for systems with highly restricted computational capacity, the SO-based method is preferred, whereas in systems with less limitations on the computational load, the PDD-based method is suggested.

We next set $\log{p}=-5$ dBm and plot the sum-rate against the number of quantization bits. The results are shown in {\figurename} {\ref{fig2b}} implying that for the schemes with optimized phase-shifts, the performance degradation caused by phase-shift quantization is negligible, as the resolution exceeds $4$ bits. Comparing the proposed algorithms with the random scheme highlights the significance of the phase-shift design. From {\figurename} {\ref{fig2b}}, it is further observed that the SO-based method outperforms the AO-based scheme for RISs of smaller size.

{\figurename} {\ref{fig2c}} illustrates the sum-rate versus the number of RF chains for selected BS antenna numbers. From the figure, one can observe that in all the presented cases, the increase in the number of RF chains enhances the system throughput.

\section{Conclusion}
We proposed an iterative algorithm based on the PDD method for joint antenna selection and beamforming in RIS-aided MU-MIMO systems. An alternative algorithm with reduced complexity was further developed based on the SO scheme. To reduce the complexity, we also developed a SO-based algorithm. Our numerical results imply that the PDD-based method outperforms both the SO-based and benchmark schemes. The SO-based algorithm moreover tracks closely the benchmark performance while gaining considerably in terms of complexity.

\end{document}